\documentclass[final,5p,twocolumn]{elsarticle}
\biboptions{numbers,sort&compress}
\usepackage{amssymb}
\usepackage{lipsum}
\usepackage{lineno}
\usepackage[T1]{fontenc}
\usepackage{newtx} %fonts
\usepackage{amsmath}
\usepackage{amssymb}
\usepackage{graphicx}
\usepackage{dcolumn}
\usepackage{bm}
\usepackage{braket}
\usepackage{mathrsfs}
\usepackage{booktabs}
\usepackage{multirow}
\usepackage{url}
\usepackage{physics}
\usepackage{subfigure}
\usepackage{float}
\usepackage{flushend}
\usepackage[colorlinks=true,linkcolor=blue, filecolor=blue,urlcolor=blue,citecolor=blue]{hyperref}
\usepackage{color}
\journal{Physics Letters B}

\begin{document}

\begin{frontmatter}

\title{Nuclear $\alpha$-cluster structures from valence-space microscopic cluster model}

\author[first]{Zhen Wang}
\ead{wang\_zhen@tongji.edu.cn}
\author[second]{Dong Bai}
\author[first,third]{Zhongzhou Ren\corref{cor1}}
\ead{zren@tongji.edu.cn}

\affiliation[first]{organization={School of Physics Science and Engineering},%Department and Organization
	addressline={Tongji University}, 
	city={Shanghai},
	postcode={200092}, 
	country={China}}

\affiliation[second]{organization={College of Mechanics and Engineering Science},%Department and Organization
	addressline={Hohai University}, 
	city={Nanjing},
	postcode={211100}, 
	country={China}}  
\affiliation[third]{organization={Key Laboratory of Advanced Micro-Structure Materials},%Department and Organization
	addressline={Tongji University}, 
	city={Shanghai},
	postcode={200092}, 
	country={China}}

\begin{abstract}
Alpha clustering is an important dynamic in nuclear physics, with growing interest to its study in heavy nuclei in recent years. Theoretically, the microscopic cluster models taking nucleons as relevant degrees of freedom have been widely used to study $\alpha$-cluster structures in light nuclei. However, a straightforward application on same footing in heavy nuclei is obstructed by the complexity of handling numerous nucleons. As a simplified alternative,  the macroscopic cluster models built upon cluster degrees of freedom are usually employed in heavy nuclei, though these approaches typically lose several critical structural details.  In this work, we propose to study the $\alpha$-cluster structures within the framework of valence-space microscopic cluster model (VS-MCM), which is a hybrid between microscopic and macroscopic cluster models and inherits features from both models, making it capable to investigate the $\alpha$-cluster structures in heavy nuclei from a relatively microscopic viewpoint. In VS-MCM, the valence $\alpha$ clusters are described by antisymmetrized microscopic wave functions, with single-particle orbits in core nuclei removed systematically from the model space via the Pauli projection to simulate the antisymmetrization between $\alpha$ clusters and doubly magic cores. As a proof of principle, we apply the VS-MCM to study the $\alpha$-cluster structures in ${}^{20}$Ne and ${}^{44}$Ti at first, with the theoretical energy levels of the $K^{\pi}=0_1^{\pm}$ bands for ${}^{20}$Ne and ${}^{44}$Ti showing reasonable agreement with experimental data. These calculations lay the foundation for future applications of VS-MCM in general cluster structures across the nuclide chart, where more $\alpha$ clusters and valence nucleons can exist outside the heavy doubly magic core, opening new avenues to study the $\alpha$ and cluster decays in heavy nuclei microscopically.
\end{abstract}


\end{frontmatter}

%\tableofcontents

% \linenumbers

%% main text

\section{Introduction}
\label{introduction}
\begin{figure*}[t]
	\centering
	\includegraphics[width=14cm]{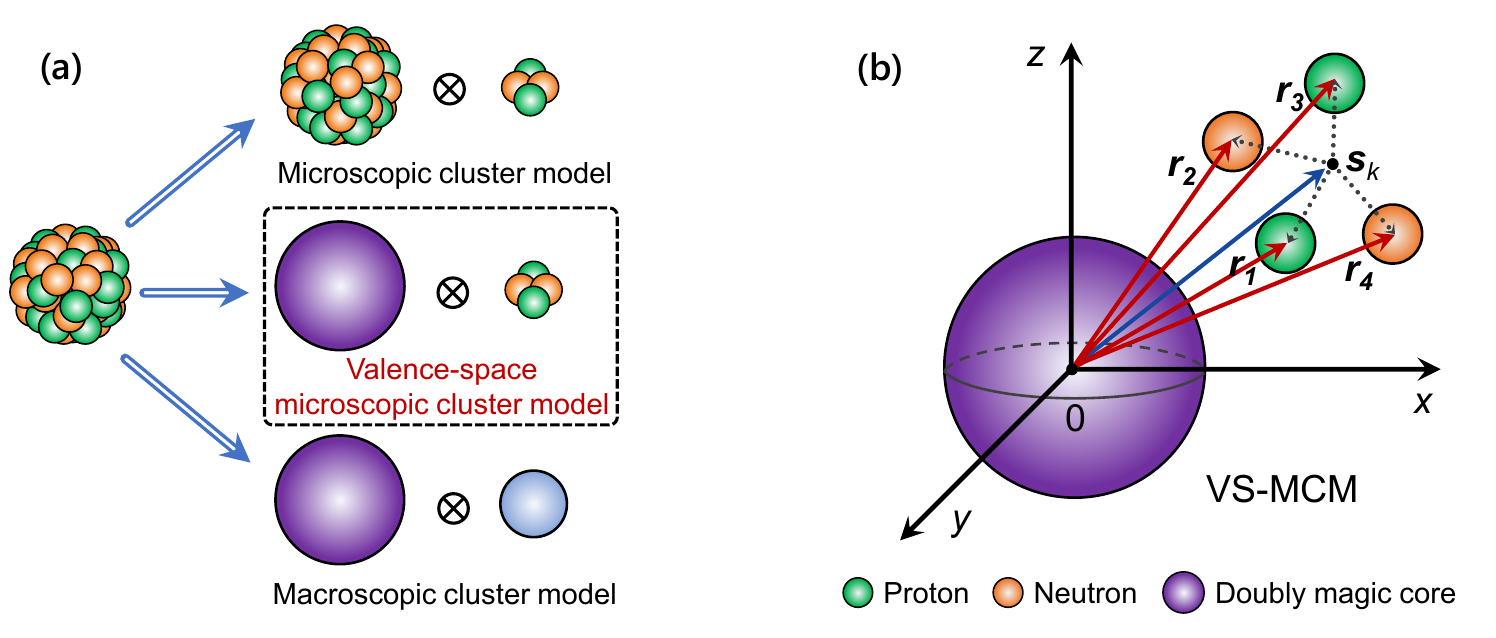}
	\caption{\label{coordinate} (a) The model spaces for $\alpha$-cluster structures above double shell closures within the framework of microscopic cluster models, valence-space microscopic cluster model (VS-MCM), and macroscopic cluster models. The microscopic structures of the $\alpha$ cluster and the core nucleus are considered to different extents in different models. See text for details. (b) The coordinate system of VS-MCM spanned from the center of mass of the core to each valence nucleon. The trial wave function for the valence $\alpha$-cluster is given by a modified Brink wave function built from four single-particle wave functions at center of $\bm{s}_{\!k}$, where the core orbits have been removed via the Pauli projection.}
\end{figure*}
Alpha clustering is ubiquitous across the nuclide chart and plays a crucial role in understanding nuclear structures~\cite{Freer:2017gip,Ren:2018xpt,Delion:2023qha,Ohkubo:2023ziz,Horiuchi:2022rdp,Taniguchi:2023dxe,Delion:2010yxh,Lombardo:2023eht,Qi:2018idv}. The existence of $\alpha$-cluster structure is well-established in light nuclei~\cite{Tohsaki:2001an,Tohsaki:2017hen,Zhou:2023vgv,Kawabata:2024vdo,Shen:2022bak,Ren:2023ued,Chernykh:2007zz}, such as $^{8}$Be, $^{12}$C, $^{16}$O, etc.  Microscopic cluster models are widely used to study their physical properties (see Refs.~\cite{Freer:2017gip,Zhou:2013ala,Kanada-Enyo:2012yif,Chien:2024jto,Tomoda:1978vlo,Neff:2010nm} and citations therein),  which take nucleons as fundamental degrees of freedom with rigorous antisymmetrization,  reliably yielding results consistent with experiments and mutually agree. As the nucleus grows heavier, the situation becomes increasingly complex. The stronger spin-orbit force and the dominant mean field tend to suppress $\alpha$ clustering~\cite{Delion:2010yxh,Dumitrescu:2022mbx}.  Nevertheless, various experimental evidences suggest the persistence of $\alpha$-cluster structures in heavy nuclei~\cite{Taniguchi:2021ulj,Zago:2022pew,Astier:2009bs,vonTresckow:2021dgi,Yoshida:2021mwf,Auranen:2018usv,Clark:2020bum,Tanaka:2021oll}. For instance, the ground state of $^{212}$Po with large $\alpha$-reduced width is proposed to have the mixing shell model configuration and $\alpha \otimes {}^{208}\text{Pb}$ configuration~\cite{Varga:1992zz}. The enhanced $E1$ transitions from non-natural parity states at excitation energies above 1.7 MeV also indicate significant alpha-cluster components in these states~\cite{Zago:2022pew,Astier:2009bs,vonTresckow:2021dgi,Yoshida:2021mwf}. Similarly, the $\alpha\otimes{}^{100}\text{Sn}$ configuration is believed to play an important role in the superallowed $\alpha$ decay of new nuclide ${}^{104}$Te,  whose $\alpha$-reduced width was deduced more than a factor of 5 larger than that for $^{212}$Po~\cite{Auranen:2018usv,Clark:2020bum}. Furthermore, direct experimental evidence for $\alpha$ clustering at the surface of neutron-rich Sn isotopes has been obtained through the detection of knocked-out $\alpha$ particles in $^{x}\text{Sn}(p,p\alpha)^{x-4}\text{Cd}$ reactions~\cite{Tanaka:2021oll}. These experimental events provide opportunities to investigate the $\alpha$-cluster structures in heavy nuclei. Given the successful applications in light nuclei, it is desirable to generalize microscopic cluster models to heavy nuclei as well. Unfortunately, a naive generalization seems obstructed by an immediate difficulty in fully antisymmetrizing the cluster wave functions due to the large number of nucleons in heavy nuclei. This makes analytic and numerical calculations exponentially difficult. Generally, the microscopic descriptions are only given for several nuclei near the doubly magic core~\cite{Varga:1992zz,Varga:1992oju,Kimura:2024zbq,Yang:2021mac,Bai:2019jmv}. For simplification, the macroscopic cluster models built upon cluster degrees of freedom are usually employed in heavy nuclei~\cite{Buck:1994zz,Ohkubo:1995zz,Mohr:2006yg,Xu:2006fq,Wang:2022yaj}, with the antisymmetrization in the relative cluster wave function being implemented only approximately via, e.g., Wildermuth conditions~\cite{wildermuthunified,Xu:2006fq,Wang:2022yaj,Buck:1994zz,Mohr:2006yg}, pseudopotentials~\cite{Orlowski:1982bh}, short-range repulsive potentials~\cite{Misicu:2006du,Zheng:2024pxj,Delion:2010yxh}, etc. Compared with microscopic cluster models, macroscopic cluster models often have a lower computational cost. They can be used to study $\alpha$-cluster structures across the nuclide chart, while these approaches typically lose critical structural details. 

\par
In this work, we propose to use valence-space microscopic cluster model (VS-MCM) as a hybrid between microscopic and macroscopic cluster models to study $\alpha$-cluster structures above double shell closures, e.g., ${}^{20}\text{Ne}$, ${}^{44, 52}\text{Ti}$, ${}^{104}\text{Te}$, ${}^{212}\text{Po}$, etc. These nuclei are treated as $\alpha + \text{doubly magic core}$ systems in VS-MCM, as well as in microscopic and macroscopic cluster models~\cite{hara1992alpha,sargsyan:2024ab,Yang:2021mac,Bai:2019jmv}.  The corresponding model spaces are given pictorially in Fig.~\ref{coordinate}(a). Compared to microscopic and macroscopic cluster models, VS-MCM takes four valence nucleons outside the doubly magic core as active degrees of freedom. The four valence nucleons form an $\alpha$ cluster by the basic assumption of cluster models. The core nucleons are, on the other hand, treated as frozen. They occupy the shell-model orbits beneath the major shell gap one by one. For simplicity, we ignore the internal excitation of the core nucleus. The valence $\alpha$ cluster is described by the antisymmetrized microscopic cluster wave function, with the core orbits removed systematically from the model space via the Pauli projection to simulate the antisymmetrization between the $\alpha$ cluster and the core nucleus. In VS-MCM, it is the nucleon-nucleon and nucleon-core potentials that are needed in the Hamiltonian. They are generally better understood than $\alpha$-core potentials. Meanwhile, the antisymmetrization between the $\alpha$ cluster and the core nucleus is realized in a similar way to nuclear shell models. This approach is more favored from the microscopic viewpoint than the rough approximations in macroscopic cluster models. The basic idea of VS-MCM was first explored in Ref.~\cite{Varga:1992zz,Varga:1992oju} as part of the cluster-configuration shell model applied to $\alpha$ decay in the ground state of ${}^{212}$Po. It hints that VS-MCM has the novel potential to probe $\alpha$-cluster structures outside heavy doubly magic cores, while preserving some essential features of microscopic cluster models. This provides an important motivation for our work. However, the development of VS-MCM was paused thereafter, leaving various pending problems.  Is VS-MCM also competitive for $\alpha$-cluster structures in lighter nuclei such as ${}^{20}$Ne and ${}^{44}$Ti?	How to implement the Pauli projection in a more realistic and systematic way?  Is it possible to use VS-MCM to study multi-$\alpha$-cluster structures outside doubly magic cores? These questions stimulate our interest in VS-MCM.
\par
As a proof of principle, we first study $\alpha$-cluster structures in ${}^{20}$Ne and ${}^{44}$Ti in present work, within the framework of VS-MCM. It is well-established that the $K^\pi\!=\!0_1^\pm$  bands in ${}^{20}$Ne could be understood by the $\alpha\otimes {}^{16}\text{O}$ structure~\cite{Tomoda:1978vlo,Yamaguchi:2023mya}. The similar picture is also applicable to ${}^{44}$Ti, where the $\alpha\otimes{}^{40}\text{Ca}$ structure plays an important role~\cite{Kimura:2004ez}. Compared to heavy doubly magic nuclei like ${}^{48}$Ca, ${}^{100}$Sn, and ${}^{208}$Pb, the core nuclei ${}^{16}$O and ${}^{40}$Ca have simpler shell-model structures. Especially, their single-particle orbits can be well approximated by harmonic-oscillator shell model, which simplifies the analytic and numerical calculations. Also, various methods have been used to study $\alpha$-cluster structures in ${}^{20}$Ne and ${}^{44}$Ti, which provide rich theoretical data to examine the reliability of VS-MCM. The following parts are organized as follows: In Sect.~\ref{sec-2}, 
the theoretical formalism of VS-MCM is given in detail. Much attention is paid to how to remove core orbits from the model space.
In Sect.~\ref{sec-3}, the numerical results are reported for ${}^{20}$Ne and ${}^{44}$Ti. We calculate the energy spectra of the $K^{\pi}=0_1^{\pm}$ bands for ${}^{20}$Ne and ${}^{44}$Ti, and compare the theoretical results with the other microscopic cluster models. In Sect.~\ref{sec-4}, the conclusions and remarks are given. The present work lays the foundation for our future projects on $\alpha$-cluster structures in ${}^{52}$Ti, ${}^{104}$Te, ${}^{212}$Po, etc, within the framework of VS-MCM. Furthermore, our method is also applicable to study $\alpha$-cluster structures in other heavy nuclei, where more $\alpha$ clusters and valence nucleons can exist outside the heavy doubly magic core.

\section{\label{sec-2}Theoretical framework}
\subsection{\label{sec-2a}Microscopic Hamiltonian in VS-MCM}

In VS-MCM, the four valence nucleons $\{p\!\uparrow,p\!\downarrow,n\!\uparrow,n\!\downarrow\}$ outside the doubly magic core are taken as the active degrees of freedom. The five-body Hamiltonian for the $\text{core}+2p+2n$ system is given in the VS-MCM coordinate system spanned from the center of mass of the core nucleus to each valence nucleon (see Fig.~\ref{coordinate}), which reads~\cite{hara1992alpha}
\begin{align}
	\hat{H}
	&=\sum_{i=1}^4\frac{\hat{\bm{p}}_{\!i}^2}{2m_i}+ \sum_{i=1}^4 \hat{U}_i + \sum_{i<j}^4\hat{V}_{i\!j}+\frac{1}{2m_c}\left(\sum_{i=1}^4\hat{\bm{p}}_{\!i}\right)^2, \label{hval1}\\
	&=\sum_{i=1}^4\left(\frac{\hat{\bm{p}}_{\!i}^2}{2\mu_i}+ \hat{U}_i\right)  + \sum_{i<j}^4\left(\hat{V}_{i\!j}+\frac{\hat{\bm{p}}_{\!i}\!\cdot\!\hat{\bm{p}}_{\!j}}{m_c}\right). \label{hval2}
\end{align}
Here ${\hat{\bm{p}}_{\!i}^2}/{(2\mu_i)}$ represents the kinetic energy of the $i$-th valence nucleon with $\mu_i$ as its reduced mass relative to the core. $\hat{U}_i$ is the effective core-nucleon potential, and $\hat{V}_{i\!j}$ is the residual interaction between two valence nucleons.  The last term in Eq.~\eqref{hval1} is sourced from the recoil kinetic energy of the core nucleus, which can be neglected when the core nucleus is significantly heavier than the $\alpha$ cluster.

\par
The effective core-nucleon potential $U(r)$ contains the nuclear part and the Coulomb one for the protons. For the nuclear part, we take 
\begin{align}
	U_{\!N}(r)\!=\!\frac{V_0}{1\!+\!\exp[(r\!-\!R)/a]}+\frac{1}{r}\frac{\mathrm{d}}{\mathrm{d}r}\frac{V_{LS}\cdot(\bm{l}\cdot\bm{\sigma})}{1\!+\!\exp[(r\!-\!R)/a]}.\label{UN}
\end{align} 
For the Coulomb part, we take
\begin{align}
	U_C(r)=\left(m_{\tau}+\frac{1}{2}\right){Z_ce^2}\cdot
	\begin{cases}
		\frac{3-(r/R)^2}{2R}, \qquad & r\leq R,\\
		\frac{1}{r}, \qquad & r>R.
	\end{cases}
	\label{UC}
\end{align}
In Eqs.~\eqref{UN} and \eqref{UC}, $V_0$ and $V_{LS}=\lambda V_0$ are the strengths of the central and spin-orbit potentials, $R=r_0A^{1/3}$ is the radius parameter, $a$ is the surface diffuseness parameter,  $Z_c$ is the charge of the doubly magic core, and $m_{\tau}$ is the isospin quantum number of the valence nucleon with $m_{\tau}\!=\!+1/2$ for proton and $m_{\tau}\!=\!-1/2$ for neutron.

\par
The nucleon-nucleon interaction $V_{ij}$ includes the nuclear part $V_{ij}^{N}$ and Coulomb one $V_{ij}^{C}$ as well, which takes the form as~\cite{Zhang:2022rfa,Bai:2020hmz}
\begin{align}
V_{ij}^{N}\!=\!\sum_{k=1}^{N_g}\!V_k\!\exp(-\frac{\bm{r}_{ij}^2}{a_k^2})(W_k\!-\!M_kP_{ij}^\sigma P^\tau_{ij}\!+\!B_k P_{ij}^\sigma\!-\!H_k P^\tau_{ij}),
\end{align}
and
\begin{align}
	V_{ij}^{C}=\left(m_{\tau_i}+\frac{1}{2}\right)\left(m_{\tau_j}+\frac{1}{2}\right)\frac{e^2}{\bm{r}_{ij}},
\end{align}
with $P^\sigma_{ij}\!\equiv\!(1+\bm{\sigma}_i\cdot\bm{\sigma}_j)/2$ and $P^\tau_{ij}\!\equiv\!(1+\bm{\tau}_i\cdot\bm{\tau}_j)/2$ being the spin and isospin exchange operators, and $N_g$, $V_k$, $a_k$, $W_k$, $M_k$, $B_k$, and $H_k$ being the parameters in the effective nucleon-nucleon potential. 

\subsection{\label{sec-2b}The modified Brink wave function}
In VS-MCM, the trial wave function for the valence $2p+2n$ system is given by a modified Brink wave function $\varPhi_{\bm{s}_{\!k}}$ built from four single-particle wave functions at center of $\bm{s}_{\!k}$ as 
\begin{align}
	\varPhi_{\bm{s}_{\!k}}=
\frac{1}{\sqrt{4!}}\,\text{det}\!\left\{
\hat{\psi}_{\!1\bm{s}_{\!k}}\!(\bm{r}_{1})\chi_{\sigma_1\tau_1}\,
\cdots
\hat{\psi}_{\!4\bm{s}_{\!k}}\!(\bm{r}_{4})\chi_{\sigma_4\tau_4}
\right\}.\label{phi_k}
\end{align}
Here $\hat{\psi}_{\!i\bm{s}_{\!k}}\!(\bm{r}_{i})$ is the spatial component of the single-particle wave function for $i$-th nucleon in $\alpha$ cluster, $\chi_{\sigma_i\tau_i}$ is the corresponding spin-isospin wave function, and $\bm{s}_{\!k}$ is the generator coordinate which could be loosely identified as the intercluster separation between the $\alpha$ cluster and core nucleus.
\par
Generally, the mean-field potential $U(r)$
is deep enough to support not only the valence orbits but also the core orbits which have already been occupied by the core nucleus. Therefore, the core orbits have to be removed from the model space. Different from the orthogonality projection method adopted in Ref.~\cite{Horiuchi:2014yua,Watanabe:2024ydu}, where a pseudopotential has been added to the Hamiltonian. In VS-MCM, this is done by applying the Pauli projection operator $\hat{\mathscr{P}}$ in the single-particle wave function $\hat{\psi}_{\!i\bm{s}_{\!k}}\!(\bm{r}_{i})$ as~\cite{beck2012clusters}
\begin{align}
		\psi_{\!i\bm{s}_{\!k}}\!(\bm{r}_{\!i})&=(\alpha/\pi)^{3/4}\exp\!\left[-\alpha(\bm{r}_{\!i}-\bm{s}_{\!k})^2/2\right],\label{wf} \\
	\hat{\psi}_{\!i\bm{s}_{\!k}}\!(\bm{r}_{\!i})&=\hat{\mathscr{P}}\psi_{\!i\bm{s}_{\!k}}\!(\bm{r}_{\!i})=\Big[1-\sum_{\gamma\in\text{core}}\!\ket{\varphi_{\gamma}}\bra{\varphi_{\gamma}}\Big]\psi_{\!i\bm{s}_{\!k}}\!(\bm{r}_{\!i}),\notag\\
	&=\psi_{\!i\bm{s}_{\!k}}\!(\bm{r}_{\!i})-\sum_{\gamma\in\text{core}}\!\braket{\varphi_{\gamma}}{\psi_{\!i\bm{s}_{\!k}}}\varphi_{\gamma}(\bm{r}_{\!i}).\label{wf_p}
\end{align}
In Eqs.~\eqref{wf} - \eqref{wf_p}, $\psi_{\!i\bm{s}_{\!k}}\!(\bm{r}_{\!i})$ denotes the single-particle orbit in the $\alpha$ particle in free space. $\varphi_{\gamma}(\bm{r}_{\!i})$ is the single-particle orbit occupied by the core nucleus, with $\gamma$ characterizing the quantum numbers. Here we denote these core orbits by approximately using the harmonic-oscillator shell model as
\begin{align}
	\varphi_{\gamma}(\bm{r})&=\varphi_{n_x}\!(x)\varphi_{n_y}\!(y)\varphi_{n_z}\!(z),\quad \gamma=\{n_x,n_y,n_z\},\\
	\varphi_n(x)&=(\beta/\pi)^{3/4}\big/\sqrt{2^{n}n!}\exp(-\beta x^2/2)H_n\Big(\sqrt{\beta}x\Big),
\end{align}
with $H_n(x)$ being Hermite polynomial of degree $n$. Explicitly, the core orbits satisfy $(n_x,n_y,n_z)=(0,0,0)$, $(1,0,0)$, $(0,1,0)$, and $(0,0,1)$ for ${}^{16}$O
and $(n_x,n_y,n_z)\!=\!(0,0,0)$, $(1,0,0)$, $(0,1,0)$, $(0,0,1)$, $(1,1,0)$, $(1,0,1)$, $(0,1,1)$, $(2,0,0)$, $(0,2,0)$, and $(0,0,2)$ for ${}^{40}$Ca.

\begin{figure*}[t]
	\centering
	\subfigure{
		\includegraphics[width=7.5cm]{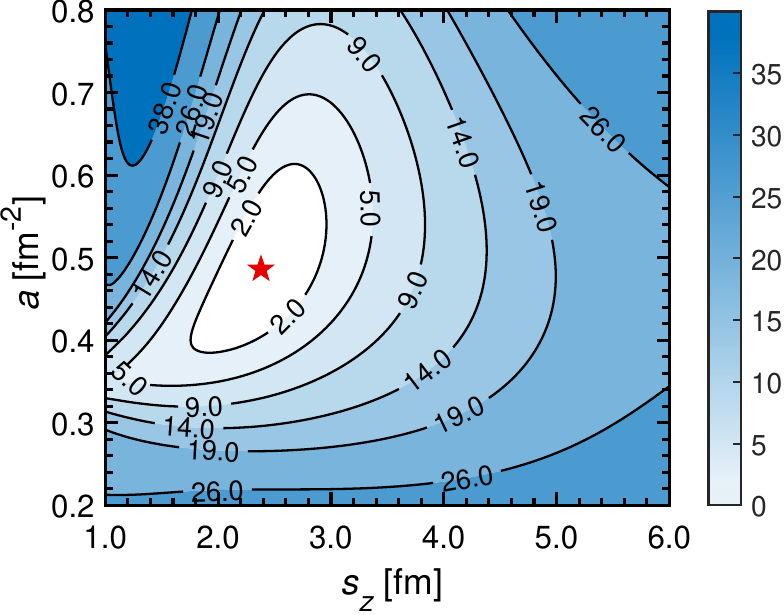}
	}\hspace{10mm}
	\subfigure{
		\includegraphics[width=7.5cm]{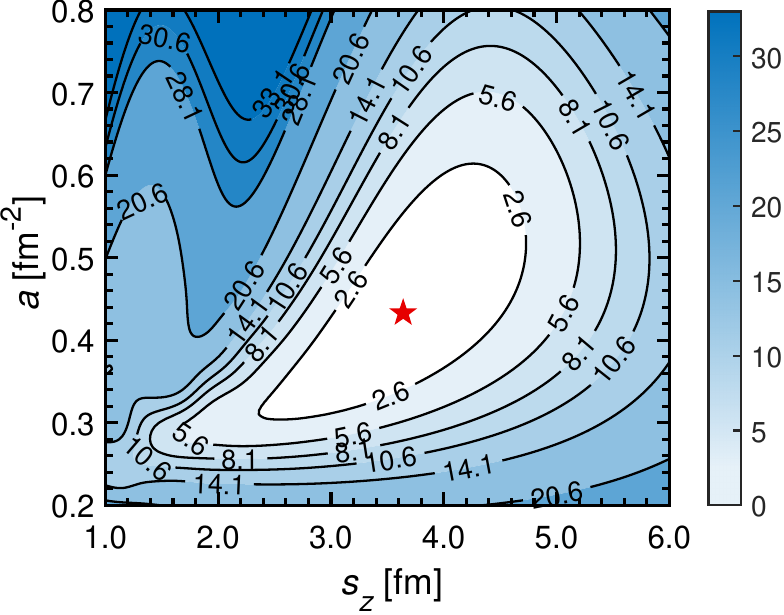}
	}
	\caption{\label{contour} Contour map of the relative energy surface $\Delta E_0(\alpha,s_z)$ for the ground state of (a) $^{20}$Ne and (b) $^{44}$Ti, in the two-parameter space of $(\alpha,s_z)$. The red star marks the location of the minimum energy.}
\end{figure*}

\subsection{\label{sec-2c}Hill-Wheeler equation}
The wave function of $\alpha$-cluster state with the definite spin $j$  is given by superposing the projected wave functions as~\cite{beck2012clusters}
\begin{align}
	\varPhi_{\!j}=\!\!\sum_k\! f^k_j \!\cdot\! \varPhi_{\!j}^{k},\quad \varPhi_{\!j}^{k}=\frac{2j\!+\!1}{8\pi^2}\!\!\int\!\!\mathscr{D}^{j*}_{mm^{\prime}}(\Omega)\hat{\mathscr{R}}(\Omega)\varPhi_{\bm{s}_{\!k}}\mathrm{d}\Omega,
\end{align}
where $\mathscr{D}^{j*}_{mm^{\prime}}(\Omega)$ is the Wigner-D function depending on the Euler angles $\Omega$, and $\hat{\mathscr{R}}(\Omega)$ is the rotation operator to perform a rotation of the wave function. Thanks for the typical property of harmonic oscillator functions, the effect of $\hat{\mathscr{R}}(\Omega)$ is equivalent to a rotation of the generator coordinates, which can greatly simplifies the calculations.
\par
The weight coefficient $f^k_j$ and the energy $E_j$ of the $\alpha$-cluster state $\varPhi_{\!j}$ can be determined by solving the following Hill-Wheeler equation:
\begin{align}
	\sum_k\left[\big<\varPhi_{j}^{k^{\prime}}\big|\hat{H}\big|\varPhi_{j}^{k}\big>-E_j\big<\varPhi_{\!j}^{k^{\prime}}\big|\varPhi_{j}^{k}\big>\right]\cdot f^k_j=0.  \label{HW}
\end{align}
The solution to the Hill-Wheeler equation is equivalent to the diagonalization of Hamiltonian in the basis of $\varPhi_{j}^{k}$ as long as the enough generator coordinates are adopted.
\par
In Eq.~\eqref{HW}, the projected Hamiltonian kernel $\big<\varPhi_{j}^{k^{\prime}}\big|\hat{H}\big|\varPhi_{j}^{k}\big>$ and overlap kernel $\big<\varPhi_{j}^{k^{\prime}}\big|\varPhi_{j}^{k}\big>$ are respectively calculated as 
\begin{align}
	&\big<\varPhi_{j}^{k^{\prime}}\big|\hat{H}\big|\varPhi_{j}^{k}\big>\!=\!\frac{2j\!+\!1}{8\pi^2}\!\int\!\!\!\mathscr{D}^{j*}_{mm^{\prime}}(\omega)\!\big<\varPhi_{\bm{s}_{\!k^{\prime}}}\big|\hat{H}\hat{\mathscr{R}}(\omega)\big|\varPhi_{\bm{s}_{\!k}}\big>\delta_{mm^{\prime}}\mathrm{d}\omega,  \notag \\
	&\big<\varPhi_{j}^{k^{\prime}}\big|\varPhi_{j}^{k}\big>\!=\!\frac{2j\!+\!1}{8\pi^2}\!\int\!\!\!\mathscr{D}^{j*}_{mm^{\prime}}(\omega)\!\big<\varPhi_{\bm{s}_{\!k^{\prime}}}\big|\hat{\mathscr{R}}(\omega)\big|\varPhi_{\bm{s}_{\!k}}\big>\delta_{mm^{\prime}}\mathrm{d}\omega, \label{mato}
\end{align}
where $\omega$ is the relative angle between $\varPhi_{\bm{s}_{\!k}}$ and $\varPhi_{\bm{s}_{\!k^{\prime}}}$. Thanks to the axial symmetry of system, we take the generator coordinate $\bm{s}_{\!k^{\prime}}$ along $z$-axis and $\bm{s}_{\!k}$ in the $xz$ plane with a relative angle $\xi$. Then Eq.~\eqref{mato} can be further simplified as
\begin{align}
	&\big<\varPhi_{j}^{k^{\prime}}\big|\hat{H}\big|\varPhi_{j}^{k}\big>\!=\!\frac{2j\!+\!1}{2}\!\int\!\!\!\big<\varPhi_{\bm{s}_{\!k^{\prime}}}\big|\hat{H}\hat{\mathscr{R}}_y(\xi)\big|\varPhi_{\bm{s}_{\!k}}\big>P_{\!j}(\cos\xi)\sin\xi\,\mathrm{d}\xi,  \notag \\
	&\big<\varPhi_{j}^{k^{\prime}}\big|\varPhi_{j}^{k}\big>\!=\!\frac{2j\!+\!1}{2}\!\int\!\!\!\big<\varPhi_{\bm{s}_{\!k^{\prime}}}\big|\hat{\mathscr{R}}_y(\xi)\big|\varPhi_{\bm{s}_{\!k}}\big>P_{\!j}(\cos\xi)\!\sin\xi\,\mathrm{d}\xi,
\end{align}
with the replacements as follows in Eq.~\eqref{mato}: 
\begin{align}
	\mathscr{D}^{j*}_{mm^{\prime}}(\omega)\Rightarrow P_{\!j}(\cos\xi),&\quad
	\int\!\!\mathrm{d}\omega\Rightarrow4\pi^2\!\!\int\!\!\mathrm{d}\cos\xi, \\
	\hat{\mathscr{R}}(\omega)&\Rightarrow\hat{\mathscr{R}}_y(\xi).
\end{align}

\section{\label{sec-3}Numerical results and discussions}

\begin{table}[b]
	\caption{The optimized parameters $V_0$, $\lambda$, $r_0$ and $a$ in the core-nucleon mean-field potential for $^{16}\text{O}+n(p)$ and $^{40}\text{Ca}+n(p)$.}
	\label{table-1}
	\centering
	\setlength{\tabcolsep}{8pt}
	\begin{tabular}{lcccc}
		\hline\hline\\[-6pt]
		&$V_0$~(MeV)&$\lambda$~(fm$^2$)& $r_0$~(fm) & $a$~(fm)\\
		\hline\\[-6pt]
		$^{16}\text{O}+n(p)$&-50.072&0.121&1.299&0.521\\[2.5pt]
		$^{40}\text{Ca}+n(p)$&-55.576&0.560&1.213&0.767\\
		\hline
		\hline
	\end{tabular}
\end{table}
\begin{figure}[h]
	\centering
	\includegraphics[width=7cm]{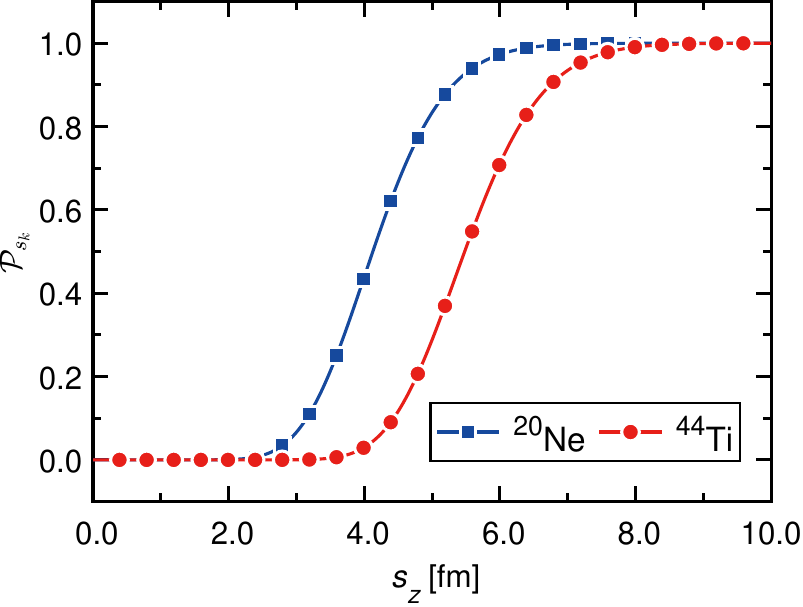}
	\caption{\label{Pa} The dependence of the Pauli index $\mathcal{P}_{\!\bm{s}_{\!k}}=\big|\big<\!\varPhi_{\bm{s}_{\!k}}(\bm{r})\big|\varPhi^0_{\bm{s}_{\!k}}\!(\bm{r})\big>\big|^2$ on the intercluster separation $s_z$, with $\varPhi_{\bm{s}_{\!k}}(\bm{r})$ and $\varPhi^0_{\bm{s}_{\!k}}\!(\bm{r})$ being the modified and standard Brink wave function respectively. The blue line with squares denote the $\mathcal{P}_{\bm{s}_{\!k}}$ for $^{20}$Ne, while the red line with circles is for $^{44}$Ti.}
\end{figure}

\begin{figure*}[t]
	\centering
	\subfigure{
		\includegraphics[width=5.85cm]{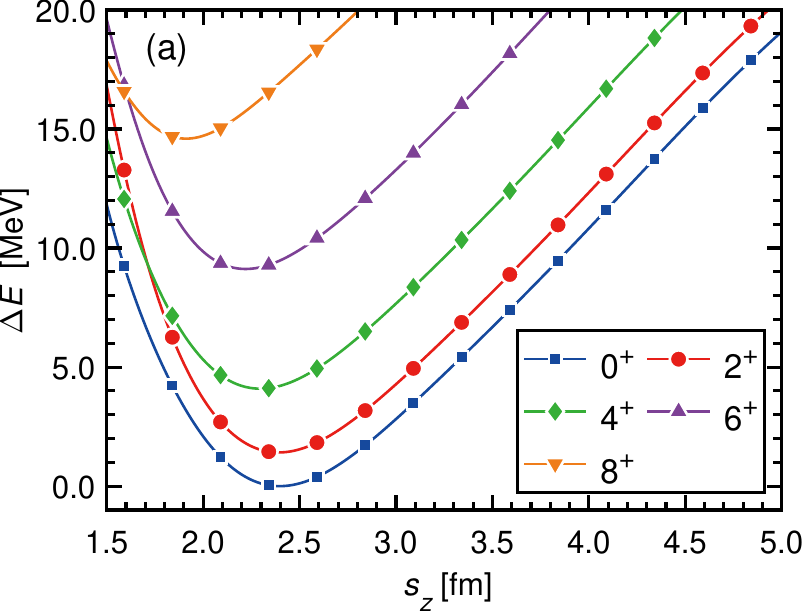}
	}%\hspace{10mm}
	\subfigure{
		\includegraphics[width=5.85cm]{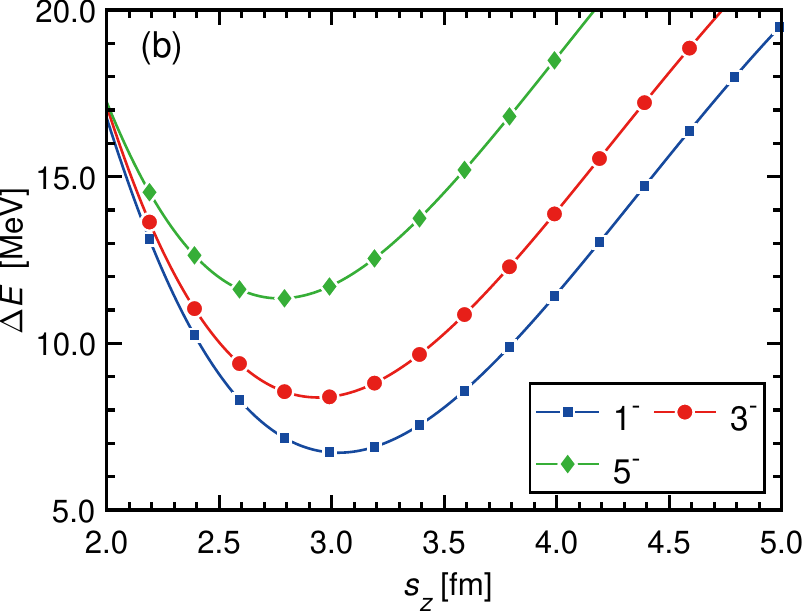}
	}
	\subfigure{
		\includegraphics[width=5.85cm]{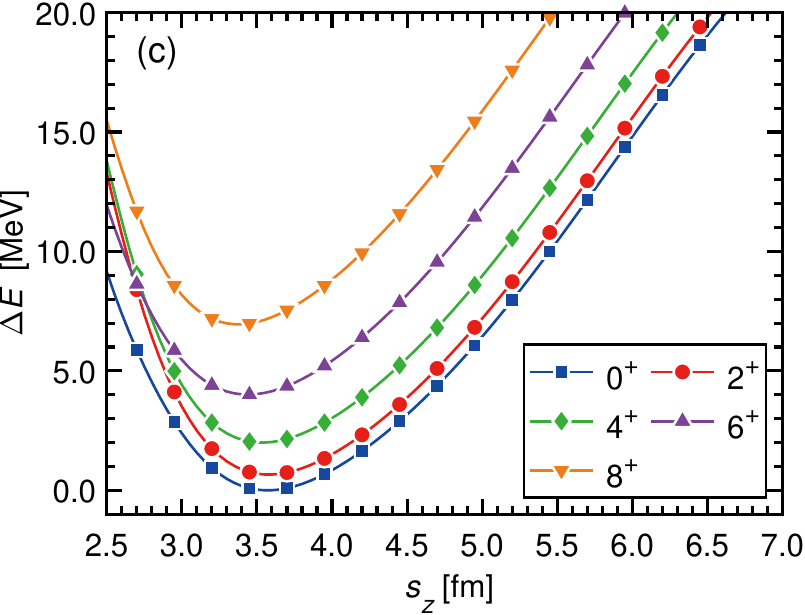}
	}
	\caption{\label{EC}The energy curves as a function of the generator coordinates $s_z$ for (a) positive band and (b) negative band of $^{20}$Ne, and (c) positive band of $^{44}$Ti. The energy are presented relative to the minimum energy of $0^+$ state in all the panels for sake of clarity.}
\end{figure*}
In this section, we apply the theoretical formalism in Sect.~\ref{sec-2} to study the $\alpha$-cluster structures  of $^{20}$Ne and $^{44}$Ti. 
We first determine the parameters in the core-nucleon potential $U(r)$ and the nucleon-nucleon interaction $V_{ij}$. For the core-nucleon parameters, we optimize their values to the experimental single-particle energies of the nuclei $^{17}\text{O}$ and $^{17}\text{F}$ for $^{16}\text{O}+n(p)$, and the energies of the nuclei $^{41}\text{Ca}$ and $^{41}\text{Sc}$ for $^{40}\text{Ca}+n(p)$ by using $\chi^2$ minimization, where the experimental single-particle energies are taken from Ref.~\cite{Schwierz:2007ve}. Here the same parameters in $U(r)$ are adopted for valence protons and neutrons for simplification. The optimized parameters are displayed in Table.~\ref{table-1}.  To facilitate the analytical calculations, we approximate the core-nucleon potential $U(r)$ by the Gaussian expansion method with $U(r)\approx\sum_kc_{\!k}\cdot\exp(-\nu_{\!k}r^2)$. The coefficients $c_{\!k}$ and $\nu_{\!k}$ are determined by the least-squares fits. For the nucleon-nucleon interaction $V_{ij}$, we take the Volkov No.~1 nucleon-nucleon interaction with the parameters $V_k$ and $a_k$ being $V_1=-83.34$~MeV, $V_2=144.86$~MeV, and $a_1=1.60$~fm, $a_2=0.82$~fm. The parameters $\{W_k,M_k,B_k,H_k\}$ are taken as $\{1-M,M,0,0\}$, where the Majorana exchange parameter $M$ can be taken arbitrarily in present work, since the energy of a single $\alpha$ cluster is independent of its value.

\par
There are two width parameters in our trial wave function, i.e., $\alpha$ and $\beta$ for $\alpha$-cluster and core orbits, to be determined as well. In present work, we fix the parameter $\beta$ in the core-orbit wave functions by reproducing the experimental root-mean-square (RMS) charge radii in the harmonic oscillator shell model. Compared with the experimental RMS charge radii $r(^{16}\text{O})=2.6991$~fm and $r(^{40}\text{Ca})=3.4776$~fm~\cite{Angeli:2013epw}, the resulting width parameter $\beta$ are found  to be $\beta(^{16}\text{O})=0.3391\,\text{fm}^{-2}$ and $\beta(^{40}\text{Ca})=0.2662\,\text{fm}^{-2}$, respectively. Within the given $\beta$ parameters, then we determine the values of parameter $\alpha$ by a full variational calculation after spin projection for the ground states in $^{20}$Ne and $^{44}$Ti.  Here we temporarily use a single projected modified Brink wave function to carry out the variational calculation in the $(\alpha, s_z)$ space, with  $s_z$ being the generator coordinates along the $z$-axis.  The energy is given by 
\begin{align}
	E_j(\alpha,s_z)=\frac{\big<\varPhi_{j}(\alpha,s_z)\big|\hat{H}\big|\varPhi_{j}(\alpha,s_z)\big>}{\big<\varPhi_{\!j}(\alpha,s_z)\big|\varPhi_{j}(\alpha,s_z)\big>}.
\end{align}
Figure~\ref{contour}(a) presents the contour map of the relative energy surface $\Delta E_0(\alpha,s_z)\!=\!E_0(\alpha,s_z)\!-\!E_0^{\text{min}}(\alpha,s_z)$ for the ground $0^+$ state of $^{20}$Ne in the $(\alpha, s_z)$ space, where $E_0^{\text{min}}(\alpha,s_z)$ corresponds to the minimum energy. As shown, the minimum energy appears at $\alpha(^{20}\text{Ne})=0.486\,\text{fm}^{-2}$ and $s_z=2.38\,\text{fm}$.  The width parameter $\alpha(^{20}\text{Ne})=0.486\,\text{fm}^{-2}$ corresponds to the cluster size of $b=1/\sqrt{\alpha}=1.43\,\text{fm}$, which is very closed to the $b$ value obtained in the nonlocalized cluster model with $b=1.46\,\text{fm}$~\cite{Zhou:2013ala}. Similarly, we also show the contour map of the relative energy surface for the ground $0^+$ state of $^{44}$Ti in Fig.~\ref{contour}(b). the minimum energy appears at $\alpha(^{44}\text{Ti})=0.433\,\text{fm}^{-2}$ and $s_z=3.64\,\text{fm}$. In the subsequent calculations, we will fix the parameters $\alpha\!=\!0.486\,\text{fm}^{-2}$, $\beta\!=\!0.3391\,\text{fm}^{-2}$ for $^{20}$Ne, and $\alpha\!=\!0.433\,\text{fm}^{-2}$, $\beta\!=\!0.2662\,\text{fm}^{-2}$ for $^{44}$Ti, respectively.

\par
Before carrying out the theoretical studies on the $\alpha$-cluster structures of $^{20}$Ne and $^{44}$Ti,  we study the properties of the modified Brink wave function introduced in Sect.~\ref{sec-2b}, within the above determined parameters.  Compared with the standard Brink wave function, the modified Brink wave function is featured by the fact that all the single-particle orbits, not only the ones belonging to the core nucleus but also those belonging to the $\alpha$ cluster, are orthogonal to each other, which makes it more convenient to interpret the wavefunction physically. Under the assumption of the inert core nucleus, the modified Brink wave function would be a better choice to describe physical properties of the $\alpha$ cluster, as it explicitly accounts for the Pauli blocking effects of the core nucleus. 
We introduce further the Pauli index $\mathcal{P}_{\!\bm{s}_{\!k}}=\big|\big<\!\varPhi_{\bm{s}_{\!k}}(\bm{r})\big|\varPhi^0_{\bm{s}_{\!k}}\!(\bm{r})\big>\big|^2$ to show the degree of Pauli blocking effects, with $\varPhi_{\bm{s}_{\!k}}(\bm{r})$ and $\varPhi^0_{\bm{s}_{\!k}}\!(\bm{r})$ being the modified and standard Brink wave function respectively.
The dependence of Pauli index $\mathcal{P}_{\!\bm{s}_{\!k}}$ on the intercluster separation $s_z$ for $^{20}$Ne and $^{44}$Ti is plotted in Fig.~\ref{Pa}. Taking $^{20}$Ne as an instance, when the $\alpha$ cluster is far away from the core nucleus $^{16}$O ($s_z>6$~fm ), the Pauli index $\mathcal{P}_{\!\bm{s}_{\!k}}\approx1$, which means that the wave function of $\alpha$ cluster is identified with that of the $\alpha$ particle in the free space. As the $\alpha$ cluster gets closer to the center of mass of ${}^{16}$O ($s_z<6$~fm), the Pauli index $\mathcal{P}_{\!\bm{s}_{\!k}}$ decreases quickly and approximately reach zero at $s_z<3$~fm, indicating the wave function of $\alpha$ cluster is gradually distorted by the Pauli blocking.  Similar trend can also be found in the ${}^{44}\text{Ti}\!=\!\alpha\otimes{}^{40}\text{Ca}$ system. The Pauli index  $\mathcal{P}_{\!\bm{s}_{\!k}}\approx1$ when $s_z>8$~fm while decreases quickly to zero when $s_z<4$~fm. Besides, the value of Pauli index $\mathcal{P}_{\!\bm{s}_{\!k}}$ is smaller than that in ${}^{20}\text{Ne}\!=\!\alpha\otimes{}^{16}\text{O}$ system. This indicates the Pauli blocking effect is much stronger in $^{44}$Ti than in $^{20}$Ne, since the core nucleus $^{40}$Ca has more nucleons than $^{16}$O, with more core orbits being occupied. The general trends agree with the theoretical expectation.  
 
\begin{figure*}[t]
	\centering
	\subfigure{
		\includegraphics[width=7cm]{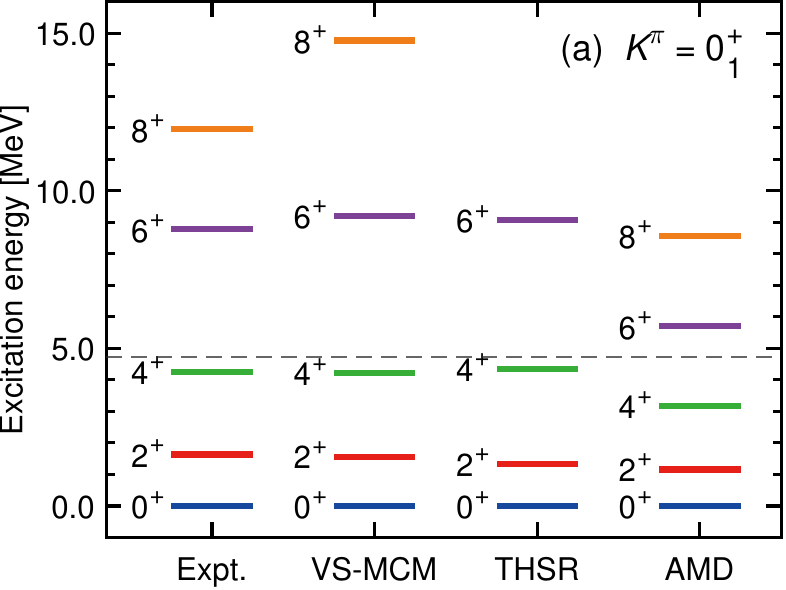}
	}\hspace{15mm}
	\subfigure{
		\includegraphics[width=7cm]{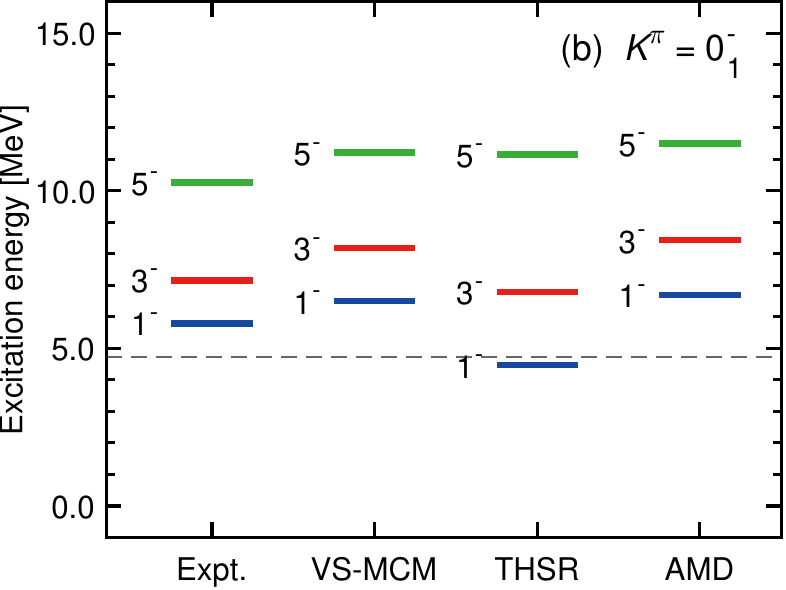}
	}
	\caption{\label{EL_20Ne}The energy levels of the inversion doublet $K^{\pi}=0_1^{\pm}$ bands  in ${}^{20}$Ne reproduced by the VS-MCM in this work, compared with the experimental data as well as the theoretical ones obtained by THSR wave function approach and AMD. The theoretical values given by THSR wave function approach are taken from Ref.~\cite{Zhou:2013ala}, while the ones given by AMD are taken from Ref.~\cite{Kimura:2003uf}. The gray dashed line indicates the  $\alpha\otimes{}^{16}\text{O}$ break threshold of 4.730~MeV.}
\end{figure*}

\par
Figure.~\ref{EC}(a) and (b) show the energy curves as a function of the generator coordinates $s_z$ for $K^\pi=0_1^+$ and $K^\pi=0_1^-$ bands of $^{20}$Ne, respectively. For clarity, the energy curves are presented relative to the minimum energy of ground $0^+$ state in both panels.  As shown in Fig.~\ref{EC}(a), the minimum energy for the ground $0^+$ state is located at $s_z=2.40$~fm, which is close to the experimental RMS charge radii of core nucleus $^{16}$O, supporting the assumption of $\alpha\otimes{}^{16}\text{O}$ configurations in $^{20}$Ne.   The minimum moves to a smaller separation when the value of $j$ increases, which reads $s_z=2.38$~fm, $2.28$~fm, $2.22$~fm and $1.91$~fm for the $2^+$, $4^+$, $6^+$ and $8^+$ states, respectively. This implies a more compact $\alpha$~+~core cluster structure in ground $0^+$ state and an undesirable overlap between core nucleus and $\alpha$ cluster in higher-spin states. Moreover, one can also find that the minimum tends to disappear for high spin states, where a broader resonance can be expected. The minimum locations of $1^-$, $3^-$ and $5^-$ states are $s_z=3.03$~fm, $2.93$~fm and $2.74$~fm, which are averagely larger than those in positive-parity band.  The larger separations obtained for the negative-parity band, on the other hand, are indicative of a more well-formed $\alpha$-cluster structure. Similar tendencies can  also be found in $K^\pi=0_1^+$ band of $^{44}$Ti in Fig.~\ref{EC}(c). The minimum of energy curve is located at $s_z=3.64$~fm for the ground $0^+$ state,  while at $s_z=3.62$~fm, $3.60$~fm, $3.53$~fm and $3.41$~fm for the $2^+$, $4^+$, $6^+$ and $8^+$ state, respectively. Different from $^{20}$Ne, all of the minimum locations of $^{44}$Ti are out of or close to the RMS charge radii of $^{40}$Ca,  implying less $\alpha$-core overlaps in these states of $^{44}$Ti.
\par
Various experimental efforts have been performed to explore the nuclear cluster structures in $^{20}$Ne and $^{44}$Ti, and the $\alpha$~+~core structure is well established in their ground bands.  We continue to study the cluster states in $^{20}$Ne and $^{44}$Ti within the framework of VS-MCM.  For $^{20}$Ne, the decay threshold is 4.730~MeV with the lowest $0^+$, $2^+$ and $4^+$ states in ground $K^{\pi}=0_1^+$ band being bound against to $\alpha$ decay. The higher $6^+$ and $8^+$ states of $K^{\pi}=0_1^+$ band, together with the lowest $1^-$, $3^-$ and $5^-$ states in negative-parity $K^{\pi}=0_1^-$ band, however, are quasibound against to $\alpha$ decay.	Besides, the $8^+$ state in $K^{\pi}=0_1^+$ band and the states in $K^{\pi}=0_1^-$ band of ${}^{44}$Ti are also quasibound above the decay threshold 5.127~MeV. Thanks to the narrow $\alpha$-decay width of these quasibound states above the $\alpha$ decay threshold, we study the energy levels of the ground bands of $^{20}$Ne and  $^{44}$Ti in the bound-state approximation. Within the aforementioned determined parameters, we calculate the energy levels by solving the Hill-Wheeler equation in Sect.~\ref{sec-2c} with spin-projected trial wave functions.  It is worth noting that the generator coordinates $\{s_z^k\}$ in the Hill-Wheeler equation should not be chosen too large and numerically close to each other, since it may cause troubles in solving the generalized eigenvalue problem~\cite{Bai:2020hmz}. In practice, one can obtain the convergent results as long as the generator coordinates include the points near the minimum of energy curves as shown in Fig.~\ref{EC}. In present work, we take $\{s_z^k\}$ as $s_z=\{0.8\,\text{fm},1.6\,\text{fm},2.4\,\text{fm},...,9.6\,\text{fm}\}$ in our subsequent calculations.

\begin{figure*}[t]
	\centering
	\subfigure{
		\includegraphics[width=7cm]{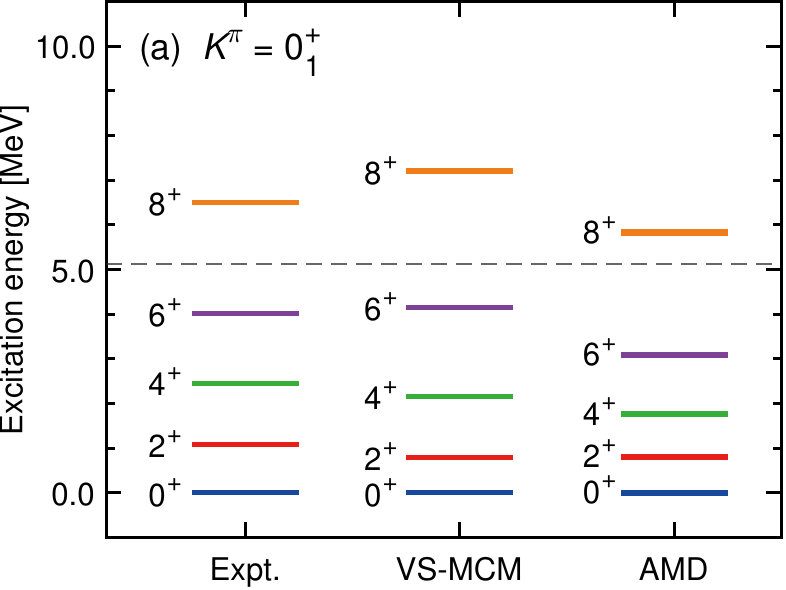}
	}\hspace{15mm}
	\subfigure{
		\includegraphics[width=7cm]{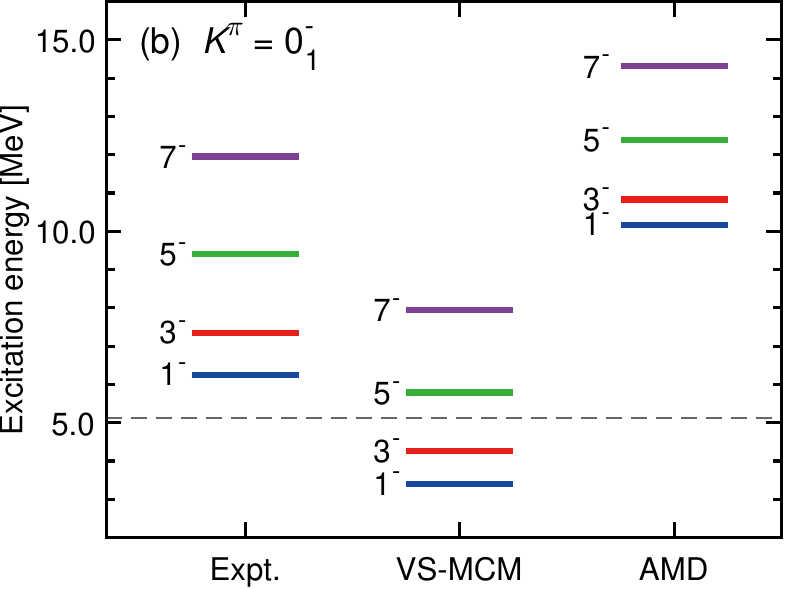}
	}
	\caption{\label{EL_44Ti}The energy levels of the inversion doublet $K^{\pi}=0_1^{\pm}$ bands in ${}^{44}$Ti reproduced by the VS-MCM in this work, compared with the theoretical ones obtained by AMD as well as the experimental data. The theoretical values given by AMD approach are taken from Ref.~\cite{Kimura:2004ez}. The gray dashed line indicates the  $\alpha\otimes{}^{44}\text{Ti}$ break threshold of 5.127~MeV.}
\end{figure*}
\par
Figure.~\ref{EL_20Ne}(a) and (b) show the energy levels of the ground $K^{\pi}=0_1^{\pm}$ bands in $^{20}$Ne reproduced by the VS-MCM. For comparison, the experimental energy levels and the ones obtained by the Tohsaki-Horiuchi-Schuck-R\"opke (THSR) wave function approach and the deformed-basis antisymmetrized molecular dynamics (AMD) calculations are also presented in both panels~\cite{Zhou:2013ala,Kimura:2003uf}. As seen, the levels in both positive and negative-parity bands are almost rotationally spaced (although the $8^+$ state is tightly compressed). Within the framework of VS-MCM, the levels of positive bands are in good agreement with experimental data, especially for the bound $2^+$ and $4^+$ states, the deviations of which are less than 30~keV.  These theoretical results are also compared with the ones given by the THSR wave function approach and AMD calculations. Nevertheless, the calculated energy of the quasibound $8^+$ state exceeds the experimental value by 2.813~MeV.  This may be due to a mixture of mean-field and cluster structure in $K^{\pi}=0_1^{+}$ band~\cite{Tomoda:1978vlo,Kimura:2003uf}. As the angular momentum increases, the $\alpha\otimes{}^{16}\text{O}$ structure component diminishes and the spin-orbit force acts strongly, which results in a reduction of $6^+$  and $8^+$ states than expected. Different from the $K^{\pi}=0_1^{+}$ band, the $K^{\pi}=0_1^{-}$ band was deduced to have an almost pure $\alpha\otimes{}^{16}\text{O}$ structure~\cite{Tomoda:1978vlo,Kimura:2003uf}, which is also supported by the VS-MCM calculations. As shown in Figure.~\ref{EL_20Ne}(b), the VS-MCM calculations give the similar spacing of $K^{\pi}=0_1^{-}$ band with the theoretical energies of $K^{\pi}=0_1^{-}$ band being slightly larger than the experimental ones by $0.718$~MeV, which is also almost identical to the theoretical results given by AMD calculations.  
\par
Compared to $^{20}$Ne, the structure of $^{44}$Ti is more complicated. The $\alpha\otimes{}^{40}\text{Ca}$ structure component in the ground band is considerably dissolved by the formation of the deformed mean-field and the spin–orbit force. Nonetheless, the AMD calculations showed that a non-small amount of $\alpha\otimes{}^{40}\text{Ca}$ structure component exists in the ground bands by about $40\%$, making it possible to use the clustering as a degree of the freedom of nuclear excitation~\cite{Kimura:2004ez}. Fig.~\ref{EL_44Ti}(a) and (b) show the theoretical energy levels of the ground $K^{\pi}=0_1^{\pm}$ bands of $^{44}$Ti, together with the experimental energy levels and the theoretical ones given by the deformed-basis AMD calculations~\cite{Kimura:2004ez}. As seen, the energy levels of $K^{\pi}=0_1^{+}$ band are reproduced very well by both the VS-MCM and AMD calculations, while only the spacing of $K^{\pi}=0_1^{-}$ band is reproduced by both models.  The bandhead of $K^{\pi}=0_1^{-}$ band obtained by VS-MCM is lower than the corresponding experimental data, this deviation may be due to the omitted mixing of mean-field structure and other cluster structures like ${}^{16}\text{O}\otimes{}^{28}\text{Si}$ configuration with the $K^{\pi}=0_1^{-}$ band~\cite{Kimura:2004ez}.  Additionally, the mixing of $^{40}$Ca core-excited state is also suggested to be very important to understand the level structure of $^{44}$Ti~\cite{Motoba:1979,Ohkubo:1998zz}, while this is beyond the framework of VS-MCM since an inert core is assumed for simplification. The present model space is not large enough to include all the above-mentioned configurations.
Addressing these mixing effects of various configurations in future studies may provide a more comprehensive understanding of the structures of $^{44}$Ti.
\par
We further calculate the root-mean-squared (RMS) energy deviation $\mathcal{S}\!=\!\sqrt{\sum_j\big(E^{\text{theo}}_j-E^{\text{expt}}_j\big)^2\big/N}$ to estimate the performance of theoretical models as presented in Table.~\ref{table-2}. The results given by VS-MCM show better or comparable agreement with the experimental data than THSR and AMD calculations, showing the reliability of VS-MCM in studying the cluster structures.

\begin{table}[hb]
	\caption{The RMS energy deviations $\mathcal{S}$ between the theoretical energy levels and experimental data for ${}^{20}$Ne and ${}^{44}$Ti.}
	\label{table-2}
	\centering
	\setlength{\tabcolsep}{4pt}
	\begin{tabular}{lcccc}
		\hline\hline\\[-6pt]
		Nucl. & $K^{\pi}$& VS-MCM (MeV)& THSR~(MeV) & AMD~(MeV)\\
		\hline\\[-6pt]
		$^{20}\text{Ne}$&$0^+$&0.245&0.248&1.897\\
		$^{20}\text{Ne}$&$0^-$&0.907&0.939&1.154\\
		\hline\\[-6pt]
		$^{44}\text{Ti}$&$0^+$&0.414&---&0.682\\
		$^{44}\text{Ti}$&$0^-$&3.418&---&3.239\\
		\hline
		\hline
	\end{tabular}
\end{table}

\section{\label{sec-4}Summary}
In this work, we propose the VS-MCM as a novel framework to investigate the nuclear $\alpha$-cluster structures especially for heavy nuclei, bridging the gap between traditional microscopic and macroscopic cluster models. VS-MCM retains key microscopic features by describing valence $\alpha$ cluster using antisymmetrized wave functions and systematically excludes core orbits via Pauli projection, enhancing computational feasibility for heavy nuclei. As a proof of principle, the VS-MCM has been applied to study the $\alpha$-cluster structures of $^{20}$Ne and $^{44}$Ti, The theoretical energy levels of $K^{\pi}=0_1^{\pm}$ bands are in  reasonable agreement with experimental data and also are compared with the ones given by microscopic THSR and AMD approaches, respectively, demonstrating the model’s reliability. 
\par
The present work lays the foundation for future applications of VS-MCM in general cluster structures across the nuclide chart, where more $\alpha$ clusters and valence nucleons can exist outside the heavy doubly magic core. It can be generalized in several ways. One direction could be to extend the modified Brink wave function to heavier clusters and core nuclei beyond the harmonic oscillator shell model adopted in the present work, in which the AMD basis can be expected. Another direction could be to relax the assumption of the inert target nucleus. This might be done by taking explicitly into consideration the nucleon exchanges between the valence clusters and the core nucleus. 
It is expected that the VS-MCM could make further improvements and bring the microscopic understanding of cluster structures in heavy nuclei in future works.

\section*{Acknowledgements}
This work is supported by the National Key R\&D Program of China (Contract No.\ 2023YFA1606503),  by the National Natural Science Foundation of China (Grants No.\ 12035011, No.\ 12447114, No.\ 12375122), and by the Postdoctoral Fellowship Program of CPSF (Grant No.\ GZB20240560).

%\bibliographystyle{elsarticle-num} 

%\bibliography{ref}

\end{document}